\documentclass[fleqn,10pt]{wlscirep}

\title{Survivability of Deterministic Dynamical Systems} 

\author[1,*,+]{Frank {Hellmann}}
\author[1,2,*,+]{Paul {Schultz}}
\author[1]{Carsten {Grabow}}
\author[1]{Jobst {Heitzig}}
\author[1,2,3,4]{J\"{u}rgen {Kurths}}

\affil[1]{Potsdam Institute for Climate Impact Research, P.O. Box 60 12 03, 14412 Potsdam, Germany}
\affil[2]{Department of Physics, Humboldt University of Berlin, Newtonstr. 15, 12489 Berlin, Germany}
\affil[3]{Institute for Complex Systems and Mathematical Biology, University of Aberdeen, Aberdeen AB24 3UE, United Kingdom}
\affil[4]{Department of Control Theory, Nizhny Novgorod State University, Gagarin Avenue 23, 606950 Nizhny Novgorod, Russia}

\affil[*]{hellmann$\|$pschultz @pik-potsdam.de}

\affil[+]{These authors contributed equally to the research presented.}

\usepackage{relsize}
\usepackage[lofdepth,lotdepth]{subfig}
\usepackage{breakurl}

\newcommand{\re}[1]{{\mathsf{Re}\left({#1}\right)}}
\newcommand{\im}[1]{{\mathsf{Im}\left({#1}\right)}}
\newcommand{\mmax}{\mathop{\max}\limits}
\newcommand{\mL}{\mathit{L}}
\newcommand{\id}{\mathbb{I}}
\newcommand{\br}[1]{\left( #1 \right)}
\newcommand{\forAll}{\displaystyle\mathop{\mathlarger{\mathlarger{\forall}}}}
\newcommand{\vol}{\mathrm{Vol}}
\newcommand{\mmin}{\mathop{\min}\limits}

\newcommand{\edit}[1]{#1}

\begin{abstract}
The notion of a part of phase space containing desired (or allowed) states of a dynamical 
system is important in a wide range of complex systems research. It has been called the safe 
operating space, the viability kernel or the sunny region. In this paper we define the notion of 
survivability: Given a random initial condition, what is the likelihood that the transient behaviour 
of a deterministic system does not leave a region of desirable states. We demonstrate the utility of this 
novel stability measure by considering models from climate science, neuronal networks and power grids.
We also show that a semi-analytic lower bound for the survivability of linear systems allows a 
numerically very efficient survivability analysis in realistic models of power grids.
Our numerical and semi-analytic work underlines that the type of stability measured by survivability is not 
captured by common asymptotic stability measures.
\end{abstract}

\begin{document}

\flushbottom
\maketitle

\begin{center}
The final publication is available at \url{http://www.nature.com/articles/srep29654} \\
or via the following DOI: \url{http://dx.doi.org/10.1038/srep29654}.
\end{center}


\section*{Introduction}

In almost all dynamical systems applicable to the real world, the stability of the system's 
stationary states (periodic orbits, chaotic attractors, etc.) is of key interest, because 
perturbations are never truly absent and initial 
data is never exactly determined. Nevertheless, the asymptotic stability of the system's attractors 
ensures that we can still extract sensible long-term information from our dynamical models. 

Complementary to the notion of stability, one can analyse whether the system will {remain in a desirable regime \cite{Heitzig2016}.}
This becomes important when a model represents a system that we have influence on, either 
because we engineer its fundamental behaviour, or because there are management 
options. We often want to design the dynamics, or our interventions, such as to more easily 
keep the system in such a desired state. {Note that the desirable region not necessarily contains a stationary state.}

For the traditional notion of asymptotic stability against {small} perturbations, the key mathematical concept is the
analysis of the linearized dynamics, in particular by means of the Lyapunov exponent or master stability function
 \cite{Nishikawa2006, Pecora1998}.
 
{ Real-world systems typically are multistable\cite{Feudel2003,Pisarchik2014,Shrimali2008}. }
 They have more than one stable attractor\cite{Milnor1985}, and thus potentially exhibit a wide range of different asymptotic 
behaviours. The key question then becomes from which initial state which attractor is reached{, i.e., to determine }
the basin of attraction of an attractor. 
Most work so far focused on the geometry of the basin of attraction \cite{McDonald1985} of desirable attractors, e.g.
by finding Lyapunov functions\cite{Belykh2004,Chiang2010,Zwillinger1997}.

A recent idea that has been found to be useful is to study a more elementary property, i.e. not which states go 
to an attractor, but just how many. 
This quantity, the volume of the basin of attraction of a given attractor, 
can then be interpreted as the {stability} of the system in the face of a random, non-small perturbation. 
It quantifies the probability that the {typically} non-linear response to such a perturbation will lead the system to a different, undesirable attractor. This probability is called the \emph{basin stability} ($S_B$) of an attractor \cite{Menck2013}.
This is important for a number of applications where relevant system deviations are typically not small, 
for example in neuro science, system Earth or power grids. 

One of the key appealing features of $S_B$ is that, by studying just the volume 
rather than the shape of the basin of attraction, it becomes numerically 
tractable to analyse even very high-dimensional systems. It was also shown that the information 
revealed by the volume of the basin genuinely complements the information provided by the Lyapunov 
exponents of the system \cite{Menck2013}. 

There are, however, two major drawbacks when estimating $S_B$.
On the one hand, the measure relies on identifying the asymptotic behaviour of a system,
which might be difficult to detect, typically requires prior knowledge about the attractor's nature,
and is only meaningful in multistable systems.
On the other hand, a $S_B$ estimation is insensitive to undesired transient behaviour of the system,
i.e. if the trajectory visits an undesired part of the phase space where the system would take damage that is not modelled {explicitly}.
To detect this type of dangerous transients, a new, complementary measure is required.

In this paper we introduce a new {stability-related measure}, the \emph{survivability}  $S(t)$ 
of a dynamical system. {This is the fraction of initial system states \edit{(i.e. arising from an initial large perturbation)} giving rise to evolutions that stay within a 
desirable regime up to a given time} $t$. The set of these initial conditions is called \emph{basin of survival}. 

More formally, call the phase space of our system $X$, and a chosen desirable region $X^+ \subseteq X$. The finite-time basin of 
survival $X^S_t \subseteq X^+$ is defined as the set of initial conditions in $X$ for which the entire trajectory 
over the interval $[0,t]$ lies in $X^+$.
We choose a probability measure $\mu$\edit{ of initital conditions,} reflecting our 
knowledge of the nature of perturbations we wish to study. 
Accordingly, the \emph{finite-time survivability} is defined as

\begin{equation}
 S_\mu(t) := \mu(X^S_t)\;.
\end{equation}

The total survivability then is the infinite-time limit of $S_\mu(t)$. This can naturally be decomposed 
into the probability that the initial perturbation is survived, and that the following trajectory stays save:

\begin{equation}
 S_\mu(t) = \frac{\mu(X^S_t)}{\mu\br{X^+}} \mu{\br{X^+}} = S_{\mu^+}(t)\cdot \mu{\br{X^+}}\;.
\end{equation}

with $\mu^+(\cdot) := {\mu(\cdot \cap X^+)}/{\mu\br{X^+}}$. Now $\mu{\br{X^+}}$ does not depend on the dynamics 
but only on the desirable region and the perturbations, i.e. it is a constant for given $X^+$. 
The conditional survivability $S_{\mu^+}(t)$ captures the interplay of dynamics, desirable region and perturbations; 
it has a natural interpretation as the conditional probability of a system to survive random, large perturbations that 
do not kill it immediately.

Assuming a uniform distribution of perturbations, the measure $\mu$ is proportional to the volume $\vol$. The resulting 
conditional survivability is our main object of study in what follows. We will call this finite-time survivability of a 
dynamical system:

\begin{equation}
 S(t) := S_{\vol^+}(t) = \frac{\vol(X^S_t)}{\vol(X^+)}\;.
\end{equation}

We are also interested in \edit{initial} perturbations that only occur in a particular region of phase space. 
Thus, we want to study uniform perturbations in a subset $C\subset X$. 
The conditional survivability $S^C\br{t}$ can then simply be defined with respect to the measure
 $\vol^C(\cdot) = {\vol(\cdot \cap C)}/{\vol\br{C}}$:

\begin{equation}\label{eq:conditional-survivability}
 S^C(t) := S_{\vol^C}(t) = \frac{\vol\br{X^S_t \cap C}}{\vol(C)}\;.
\end{equation}
 
An important example of such a conditionial survivability is the single node survivability
for networked systems. There we condition on the phase space at a single node, thereby isolating the impact of local perturbations on the whole system. A mathematically precise discussion will follow in the power grid example in the results section and the supplementary information (SI). 

\begin{figure}[!ht]
\centering
\includegraphics[width=0.4\columnwidth]{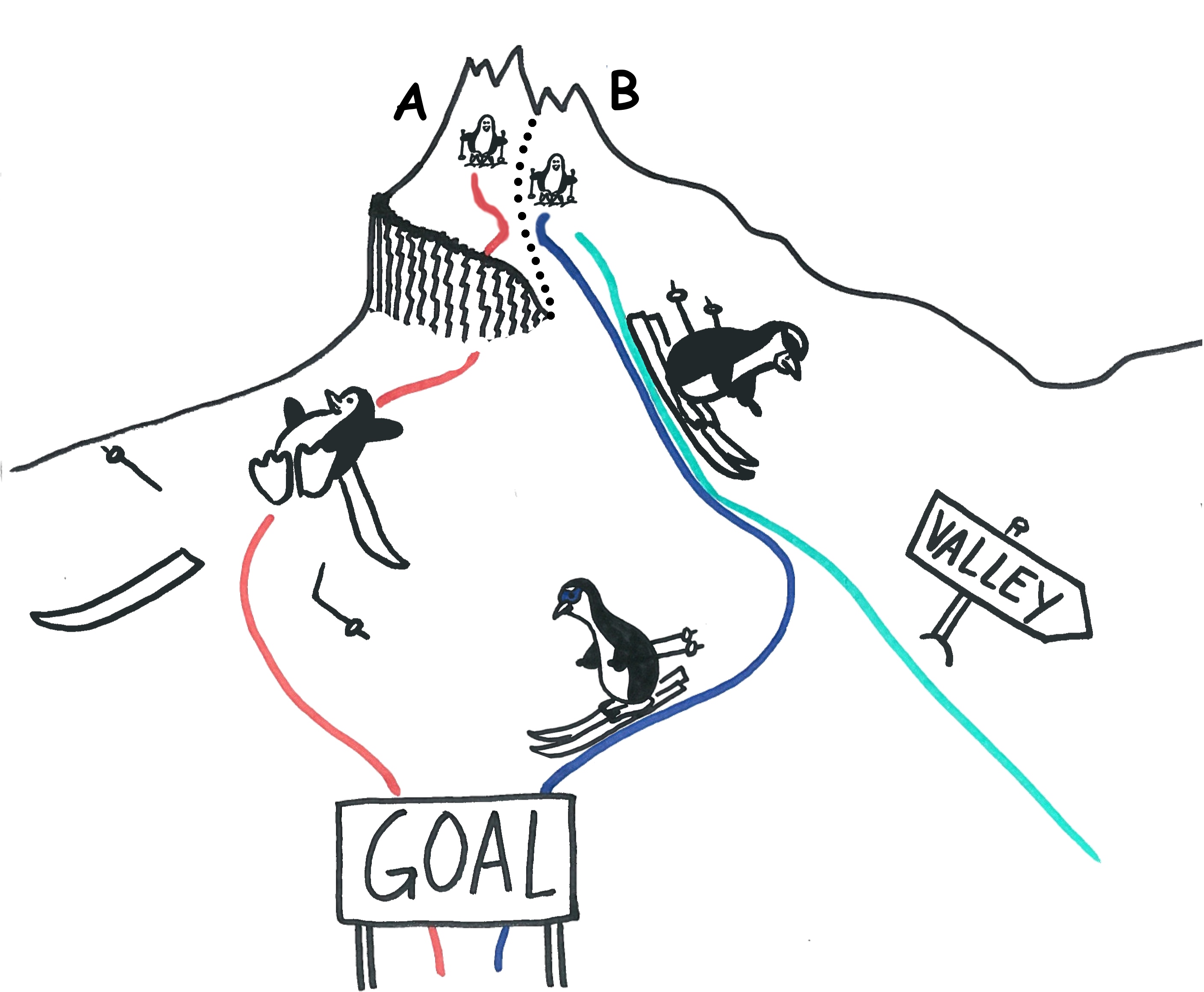}
\caption{\textbf{Survivability cartoon: } (Colour online.) A penguin can ski down the mountain starting 
anywhere on the slope. Starting at A the penguin will tumble over the cliffs, {passing an undesirable state 
although ultimately reaching the goal}. 
Starting at B the penguin will reach the goal standing on its feet. Starting even further to the right, it 
might end up in the valley, which might or might not be desirable.}
\label{fig:scheme}
\end{figure}

To further illustrate this definition, consider a simple example: A penguin wishing to ski down a 
mountain $X$ going the fastest route possible in Fig.~\ref{fig:scheme}. 
The system is multistable as the penguin might end up in the goal or the valley.
However, if the penguin goes over the cliff it will almost certainly slide the rest of the way to the goal 
on its back. The state of the penguin is not {explicitly} modelled by our (potential) landscape. 
We take this into account by declaring the parts of the cliff our penguin can not ski safely as an 
undesirable region. Further, if the penguin wishes to continue skiing, the valley might or might not 
be undesirable as well. Depending on these choices, different starting points can be in the basin of survival. 
If the goal is the only desirable attractor, the basin of survival lies in its basin of attraction, 
but if the valley is OK, too, this is not the case, and the asymptotic structure plays no role.

As opposed to $S_B$ or a linear(-ised) analysis based on Lyapunov exponents, the survivability is 
concerned not just with the asymptotic behaviour of the system, but depends strongly on the transient dynamics. 
As opposed to $S_B$ it is applicable in unstable, mono-stable, or multistable, linear or non-linear systems.

The application of the survivability concept is especially appropriate when interventions 
happen at the same time scale as the system dynamics, or when entering an undesirable region is deadly.

A key insight is that \edit{evaluating survivability becomes amenable to Monte Carlo integration. This is due to focusing on the probability that the trajectory following a perturbation violates the boundary rather than trying to find the actual sets of phase space from which a trajectory survives.} 
Hence, a survivability analysis, just as $S_B$, is applicable to very high-dimensional systems.
In fact, the situation is more favourable than in the case of $S_B$, as the entire curve $S(t)$ can be evaluated at a computational cost not exceeding that of $S_B$, while potentially revealing much more information. 

This sets survivability apart from formally similar approaches, e.g. in control theory \cite{abdallah2003necessary,amato2001finite}. Their precise relationship to survivability is discussed in detail in the \emph{Methods} section.

For linear systems with a {polyhedral} desirable region, we derive a closed form lower bound on the 
\emph{infinite-time survivability} $S_\infty := S(t\to\infty)$ as well as a semi-analytic, stronger bound that 
becomes exact in the case of {vanishing dissipation}. These bounds reveal that the survivability 
of linear systems depends strongly on the eigenvectors of the linear dynamics, rather than just 
the eigenvalues. The semi-analytic bound eliminates the need to simulate the system trajectory 
opening survivability up to a wide range of applications for which numerically estimating the full 
dynamics is not feasible.

\section*{Results}

To demonstrate the diverse applicability of our survivability concept, we apply it to three 
paradigmatic model systems. A two-dimensional model of carbon stock dynamics,
a system of integrate-and-fire neurons and a high-dimensional network model of the power grid.

These systems were chosen to cover a wide range of types of systems. The carbon cycle model has 
one or two attractors, depending on the parameter regime, and some transients are deadly. The neurons are 
mono-stable but exhibit transient chaos\cite{Houghton2010,Tel1990, Tel1996, Wolfrum2011, Lai2011}. Finally the power grid 
model is high-dimensional, non-linear and multistable. However, the acceptable operating regime is close to 
a certain class of fixed points, thus the linearized behaviour near these fixed points is of great practical importance.

In all three systems there are externalities which are not or cannot be modelled {explicitly}. Namely, 
the influence of dramatic climate changes on society, external stimuli for a network of neurons and 
frequency control mechanisms in the power grid. We will see that survivability accurately captures 
the interplay of externalities with the intrinsic dynamics.

\subsection*{Carbon cycle model by Anderies et al.}

We begin by applying survivability to a two-dimensional carbon cycle model 
from climate science which has been recently introduced \cite{Anderies2013}. This is a conceptual model with the aim 
to reproduce the non-linear dynamics of the carbon cycle in the Earth system. The boundaries of the \emph{survival region} are 
closely related to the concept of planetary boundaries \cite{Rockstroem2009}. This system exhibits both the property that 
the undesirable states are deadly and that in some parameter regimes there is only a single stable attractor of the asymptotic dynamics.

The model equations for the atmospheric ($c_a$), marine ($c_m$) and terrestrial ($c_t$) carbon stocks are given by
\begin{align}\label{eq:anderiesmodel}
\dot c_m &= \alpha_m\br{c_a -\beta c_m}   \\ \nonumber
\dot c_t &= NEP\br{c_a, c_t} - \alpha c_t  \\ \nonumber
c_a &= 1 - c_m - c_t   
\end{align}

where $\alpha_m$ denotes the atmosphere-ocean diffusion coefficient, $\beta$ the carbon-solubility in sea water factor, $\alpha$  
the human terrestrial carbon off-take rate and $NEP\br{c_a, c_t}$ the net ecosystem production, a complex non-linear relationship
between the atmospheric and terrestrial carbon stocks (see Anderies et al.\cite{Anderies2013} for further details). Note that the total amount 
of carbon is kept constant, leaving us with the marine ($c_m$) and terrestrial ($c_t$) carbon stocks as  independent variables.

Part of the phase space $X$ of the model are states with virtually no terrestrial carbon, referred to as \emph{desert states}. While the model can
recover from such states and eventually reach high terrestrial carbon states again, entering a desert state would lead to the collapse
of human civilisation and thus, tragically, our model would no longer be valid after entering this regime. Hence, we define the set of desirable
 states $X^+$ as the  complement of the desert states plus a safety margin $m$:
\begin{equation}
X^+= \{(c_a, c_m, c_t)\in X : c_t > m\} \ .
\end{equation} 

The safety margin should at no time, during the transient or asymptotic behaviour, be crossed. 
The finite-time basin of survival, here introduced as $X^S_t$, is then given by
  
\begin{equation}
X^S_t= \{(c_a, c_m, c_t)\in X : \forAll_{0 \leq t'\leq t}\, c_t(t') > m\} \ .
\end{equation} 


\begin{figure}[!ht]
\centering
\includegraphics[width=\columnwidth]{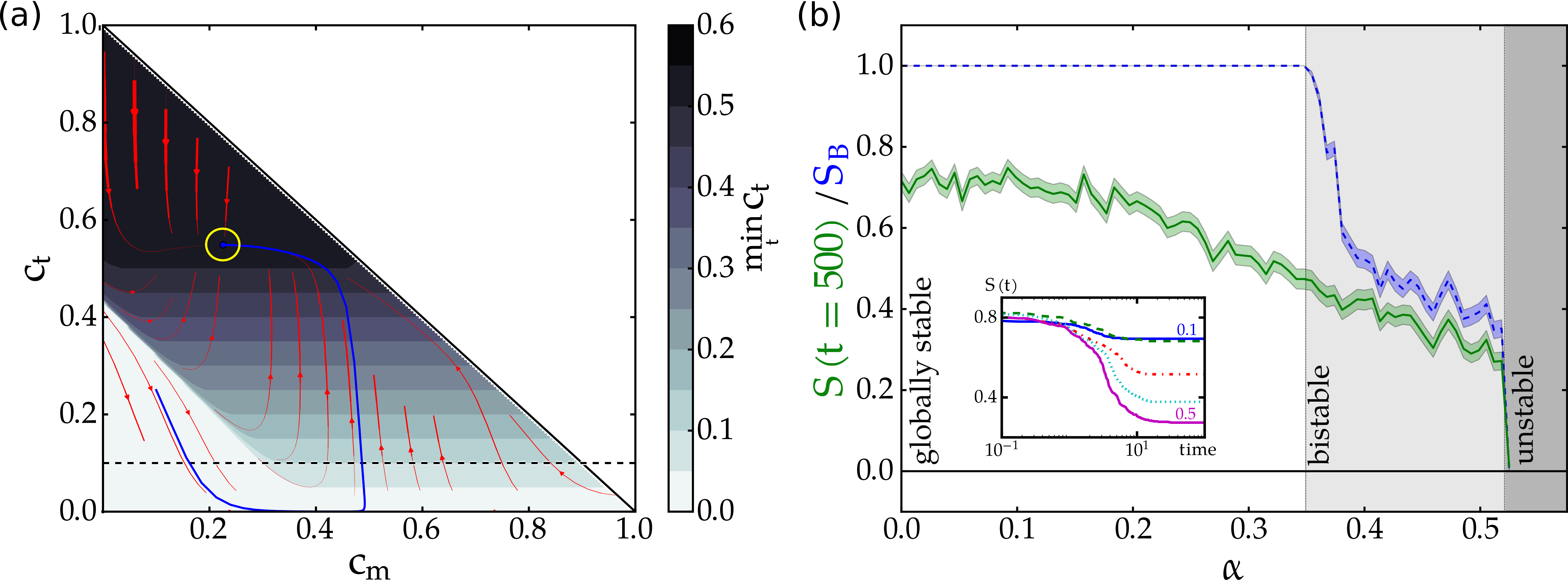}
\caption{\textbf{(a) Phase portrait of Anderies' model (Eqn.~\ref{eq:anderiesmodel}, $\alpha=0.1$).} (Colour online.) 
{We choose initial terrestrial ($c_t$) and marine ($c_m$) carbon stocks, the colour scale then indicates the 
minimum of $c_t$ over the whole time evolution commencing from a point. }
An example trajectory with a long excursion to the desert state ($c_t < m$) is plotted in blue \edit{and ends at the attractor which is circled in yellow}, the 
stream plot indicates the vector field of the right-hand-side (cf. Eqn.~\ref{eq:anderiesmodel}). 
The dashed black line indicates the value of the safety margin $m=0.1$. \\
\textbf{(b) {Bifurcations in the carbon cycle model}:} (Colour online.) 
{Basin stability ($S_B$, blue) and finite-time survivability ($S(t=500)$, green) estimates for different values
of the terrestrial human carbon off-take $\alpha$.} For the survivability estimation we assumed a safety margin $m=0.1$.
The shading around the curves indicates one standard error, the background colour indicates the different dynamical 
regimes. \edit{In the inset, we give survivability curves for five selected values of $\alpha$, i.e. 
$\alpha\in\{0.1, 0.2, 0.3, 0.4, 0.5\}$ from top to bottom as indicated.}}
\label{fig:bos-and_alpha}
\end{figure}

A phase plane analysis for this model is illustrated in Fig.~\ref{fig:bos-and_alpha}a. Of special importance here are those trajectories 
(exemplified by the blue trajectory in Fig.~\ref{fig:bos-and_alpha}a) that first cross the safety margin, i.e. are not desirable due to the very 
low terrestrial carbon stocks $c_t$, but eventually will return to the {desirable region} $X^+$. 
These trajectories are counted for the $S_B$ estimation, since they eventually approach the attractor,
but are disregarded for the survivability, since they cross the safety margin during the transient period.

By varying the human carbon off-take $\alpha$ in Eqn.~\ref{eq:anderiesmodel}, the system undergoes a bifurcation changing 
the number of attractors (around $\alpha=0.35$) as illustrated in Fig.~\ref{fig:bos-and_alpha}b. \edit{The main picture shows the asymptotic survivability, the inset contains the survivability curves for different values of $\alpha$. We see that the survivability drops to the asymptotic plateau at around the same time. Thus, if a trajectory eventually leaves the desirable regime, the time it takes until it does so is not strongly affected by $\alpha$.}

The bifurcation, which is known to be a saddle-node bifurcation \cite{Anderies2013}, has a drastic impact on the 
$S_B$ estimation, the survivability only changes marginally in this interval. On the other hand, the behaviour 
in the interval $\alpha\in\left[0;0.35\right]$ shows how the $S_B$ estimation becomes insensitive to system 
changes if the multistability is lost, i.e. if there is only a single attractor (in this case with non-zero $c_t$). The crucial 
question whether trajectories stay in a {desired} regime is thus not captured by the $S_B$ measure, but can be answered 
with the survivability concept. Note that in this case and in what follows we estimate a finite-time survivability for the entire simulated time 
evolution of the system. Given that the asymptotic behaviour sets in earlier than the simulation ends, this is a good estimate 
for the infinite-time survivability.

It was argued \cite{Menck2013} that $S_B$ can also serve as a better early warning indicator of {approaching
tipping points} than other measures. Here we see that a survivability estimation mirrors the trend in the system's behaviour, 
i.e. how the set of surviving states depends on system parameters, while $S_B$ remains fixed at its plateau value. 
Hence,  survivability can serve as a complementary, and in some scenarios better early warning sign than $S_B$. 

\subsection*{Network of integrate-and-fire oscillators}

 \begin{figure}[!ht]
\centering
\includegraphics[width=0.4\columnwidth]{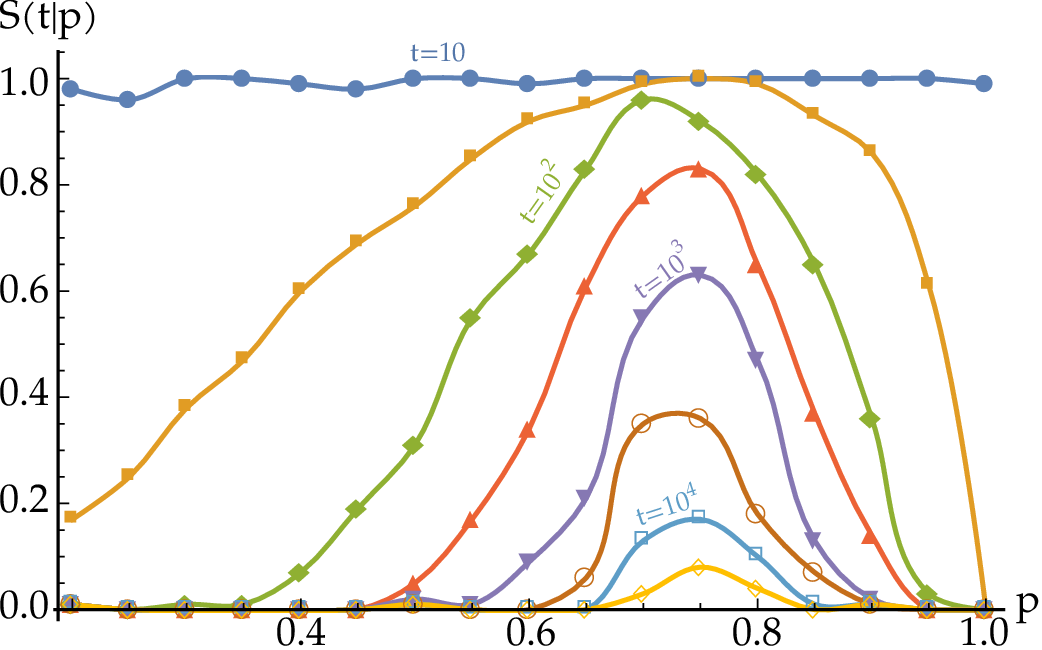}
\caption{\textbf{Survivability curves for networks of integrate-and-fire oscillators:} (Colour online.) 
Finite-time survivability $S(t|p)$ for given survival times $t$ vs. the network parameter $p$. For each 
value of $p$ we average over an ensemble of $100$ network realisations, each with \edit{initial conditions 
drawn at random from the full state space}.}
\label{fig:pcon}
\end{figure}

In the case of transient chaos\cite{Houghton2010,Tel1990, Tel1996, Wolfrum2011} there are long, 
interesting transients but potentially just a single global attractor.
As an example, we consider a network of $N$ integrate-and-fire neurons 
\cite{Ernst:1995:1570,Jahnke:2008p2847,MIROLLO:1990p366,Winfree:1614992}.
They exhibit long-term chaotic transients, but asymptotically have a global periodic 
attractor where the neurons are in a state of phase-synchronisation.
Considering the synchronised state as undesirable, the integrate-and-fire neurons are an example of a
system in which neither asymptotic nor basin stability are {informative}.

Modelling external stimuli as essentially randomly resetting the phases of stimulated neurons, 
the survivability $S(t)$ here carries the interpretation of the probability that the system will not 
fall into a synchronised state in between stimuli, spaced apart at interval $t$. {Such synchronised 
states model epileptic seizures and are thus undesired.}

Concretely we study the convergence from arbitrary initial conditions to periodic orbit attractors, 
in which several synchronised groups of oscillators (clusters) coexist \cite{Zumdieck:2004be}. 
In the network every oscillator $j=1\dots N$ is connected to another oscillator $i\neq j$ by
a directed link with probability $p$. A phase variable $\phi_{j}(t)\in[0,1]$ specifies the state of each oscillator 
$j$ at time $t$.\edit{
The free dynamics of an oscillator $j$ is given by
\begin{equation}
\dot\phi_{j}(t)=1\,.
\end{equation}

 The oscillators interact on a directed graph by sending pulses when they reach the threshold $\phi_j = 1$.
 After a delay time $\tau$ this pulse induces a phase jump (indicated by differentiating the left and right limit of $t$ as $t^+$ and $t^-$)
in the receiving oscillator $i$:

\begin{equation}
\phi_{i}(t^+):=U^{-1}(U (\phi_i(t^-) - \epsilon_{ij}))
\end{equation}

for a potential $U$ and coupling strength $\epsilon_{ij}$ (For more details cf. the {SI}).}

The survivability $S(t|p)$ for a directed network of $N=16$ pulse-coupled oscillators in dependence 
on the average connectivity $p$ is illustrated in Fig.~\ref{fig:pcon}. For each value of $p$ we create 
an ensemble of $100$ network realisations. The randomly chosen initial phase 
vectors for each realisation are distributed uniformly in $[0,1]^{N}$. 

All different network realisations 
with their associated initial conditions eventually lead to a fully synchronous state. However, our concept 
of survivability reveals the highly non-linear, non-monotonic dependence on the network connectivity $p$. 
While the survivability of transient dynamic states is small for networks with low and high connectivity values $p$,
it becomes very large for intermediate connectivities, even for only weakly diluted networks (Fig.~\ref{fig:pcon}). 
The finite-time survivability reveals a new, collective time scale that is much larger than the natural period, $1$, 
of an individual oscillator and the delay time, $\tau$, of the interactions. 

These long, irregular transients are the main property of interest for the system, motivating their study 
in \cite{Zumdieck:2004be}. The dependence of the average lifetime of the transient chaotic trajectories 
on $p$ was already studied ibidem. In this example, survivability reveals the same dynamical information 
as previous studies. Note that this is due to the specific choice of desirable region \edit{as the non-periodic parts of state space}. Generally, there is no 
direct relationship between survivability and transient lengths, \edit{the fact that the desirable region can be chosen such that survivability reveals the quantity of interest} for this system in a natural way speaks for its universality.

Survivability again is a natural and informative stability measure of this system, however, this time 
not against perturbations, but against getting trapped in an undesired corner of phase space.

\subsection*{Power grids}\label{sec:power grid application}

Power grids are subject to a variety of failures and perturbations and there
are numerous studies concerning asymptotic stability analysis, e.g. \cite{Doerfler2013,Motter2013}, 
and recent approaches to an $S_B$ assessment \cite{Menck2014,Schultz2014}. 
However, contrary to common model assumptions, the dynamical system does usually not 
evolve freely after a perturbation. If the system does not return to a stable 
operating state after a typical time span of a few seconds or
if predefined thresholds are exceeded, control mechanisms that would require independent modelling are triggered. 

The long-term behaviour and stability of the system is thus a question for control theory rather than just dynamics.
Conversely, the transient dynamics, and the question whether there is a temporary amplification of perturbations, is 
critical to whether the control has to be activated at all, or the system is {explicitly} resilient to such perturbations.
Hence, the power grid is an example where the undesirable region is deadly and 
management options operate at the system dynamics time scale.

The effective network model of the power grid \cite{Filatrella2008, Nishikawa2015}
 is the current standard baseline model for the frequency 
dynamics of power grids. It is known as the \emph{swing equation} or the second-order Kuramoto model, 
and is used for short-term frequency 
stability studies in power grids. The various ways in which a power grid can be modelled using the swing equation are discussed in 
\cite{Nishikawa2015} and limits to its applicability are discussed, for example, in \cite{Weckesser2013, Auer2016}. 

The dynamical system modelling \edit{$N$} generators' instantaneous phases $\phi_i$ and frequency deviations $\omega_i$ 
from the grid's rated frequency is given as 
\begin{align}
 \dot\phi_i &= \omega_i \\
\nonumber \dot\omega_i &= P_i - \alpha_i\omega_i - \sum\limits_{j=1}^N\mathit{K}_{ij}\sin\br{\phi_i-\phi_j}
\end{align} 
with $P_i$ being the net input power/consumption, $\alpha_i$ the electro-mechanical damping at node $i$
and $\mathit{K}_{ij}$ as the capacity of the link $i$ -- $j$. Here we choose $P_i=1$ for net generators, $P_i=-1$ for 
net consumers, and a uniform distribution of $\alpha_i = \alpha = 0.1$. We choose the nonzero $\mathit{K}_{ij}$ 
uniformly equal to $6$, corresponding to an average transmission line length of about 200 km.

A stable operating state of the power grid is a fixed point of the dynamics with no frequency deviation, 
$\br{\phi^*, 0} :=\br{\phi^*_1, \dots , 0, \dots}$. 
Conversely, limit cycle solutions (frequency oscillations) need to be prevented 
in order to avoid the tripping of generators. Frequency deviations are usually kept very small in large real power 
grids, with typical thresholds of $\pm0.2$\,Hz  \cite{UCTE2004} which corresponds to a phase velocity deviation of 
$|\omega|\approx 0.25$ in our units. Smaller island grids have considerably larger fluctuations. \edit{As an illustrative extreme case we will consider up to $20$ times larger fluctuations.}
For $S_B$ assessments, the reaction of the system to much larger deviations was also taken into account.

\edit{We will study the \emph{single-node basin of survival}, i.e., the conditional basin of survival in the sense of Eqn.~\ref{eq:conditional-survivability}, conditioned on initial perturbations that occur locally at a single node $n$, starting from a stable operating state.
The space we wish to condition on is then the direct product of the stable operating state at all nodes except node $n$ and the full state space of the node dynamics at $n$:}

\begin{align}
\nonumber C_n = \{ \br{\phi^*_1, \dots \phi_n, \dots, \phi^*_N, 0, \dots, \omega_n, \dots, 0}
| \phi_n \in [0,2\pi), \omega_n \in \mathbb{R} &\} \; .
\end{align}

The desirable region being defined as $\forall i:\,|\omega_i| < 5$, which, as explained above, is chosen to mirror realistic constraints. 
Concretely, this means that we construct initial conditions by setting $\phi_i$ and $\omega_i$ to the value of the fixed point $\phi^*_i$ and $0$, 
for all nodes other than the node $n$
we are studying, and to a random phase in $[-\pi;\pi]$ as well as a random frequency deviation in $[-5;5]$ for the node $n$.
Then we simulate the system up to $t=100$ and observe whether (and if, when) any of the frequency deviations 
$\omega_i$ \edit{leave} the desirable region. In this way we sample 300 trajectories to estimate 
$S^n(t) := S^{C_n}(t)$.
This leads to a standard error of less than $0.03$ for $S^n(t) = 0.5$ in the worst 
case (see Methods section). We evaluate the survivability up to $100$ in simulation time ($18$s in real time), 
at which point a steady state has typically been established, and the asymptotic value of the survivability is reached.

While $S_B$ captures the overall ability of the system to avoid permanent frequency oscillations, it does not
directly capture the {stability} of the system against large perturbations.  Instead, as discussed above, it is the ability of the 
system to keep perturbations under fixed frequency thresholds which is crucial. We will 
study this form of {stability} using both numerical simulations and the analytic approximations we have derived. 
The former will allow us to compare the survivability of the system to its $S_B$, the latter to assess the accuracy of our bounds.

We now turn to the question whether the semi-analytic bounds on the dynamics 
linearized around the fixed point can accurately mirror the single-node survivability $S^n(t=100)$.

Defining $\phi:=\br{\phi_1,\dots,\phi_N}^T$, $\omega=\br{\omega_1,\dots,\omega_N}^T$ and $\alpha := diag(\alpha_i)$,
the linearized dynamics is given by  

\begin{equation}
 \begin{pmatrix}
 	\dot\phi\\\dot\omega
 \end{pmatrix} = 
 \begin{pmatrix}
 	0&\id_N\\L&-\alpha\end{pmatrix}\begin{pmatrix}\phi\\\omega
 \end{pmatrix}
\end{equation}

where the lower left block ($L=\partial\dot\omega_i/\partial\phi_j$) can be identified with the network's 
Laplacian matrix (at the fixed point $(\phi^*, 0)$) given by
\begin{align}
 \mL_{ij}=- \delta_{ij}\sum\limits_{m=1}^N & \mathit{K}_{im}\cos\br{\phi_i^*-\phi_m^*}+
 \mathit{K}_{ij}\cos\br{ \phi_i^*-\phi_j^*}
\end{align}

The Jacobian has two real eigenvalues, $\lambda_1 = 0$ and $\lambda_2 = - \alpha$, corresponding to the 
eigenvectors $\br{\phi,\omega}_1= \br{1,\dots, 0,\dots}$ and $\br{\phi,\omega}_2 = \br{-1/\alpha, \dots, 1, \dots}$.
The first \edit{eigenvalue, $\lambda_1$ and the corresponding eigenvector show} the linearized version of the rotational symmetry of 
the system under shifting all elements of $\phi$ by the 
same amount $\phi_s$: $\phi_i \mapsto \phi_i + \phi_s$. The second corresponds to a homogeneous shift of all 
oscillator's frequencies, which does not affect the phase differences, and decays exponentially due to the damping term. 
The remaining part 
of the spectrum consists of $N-1$ pairs of complex conjugated eigenvalues.

\begin{figure}[!ht]
\centering
\includegraphics[width=\columnwidth]{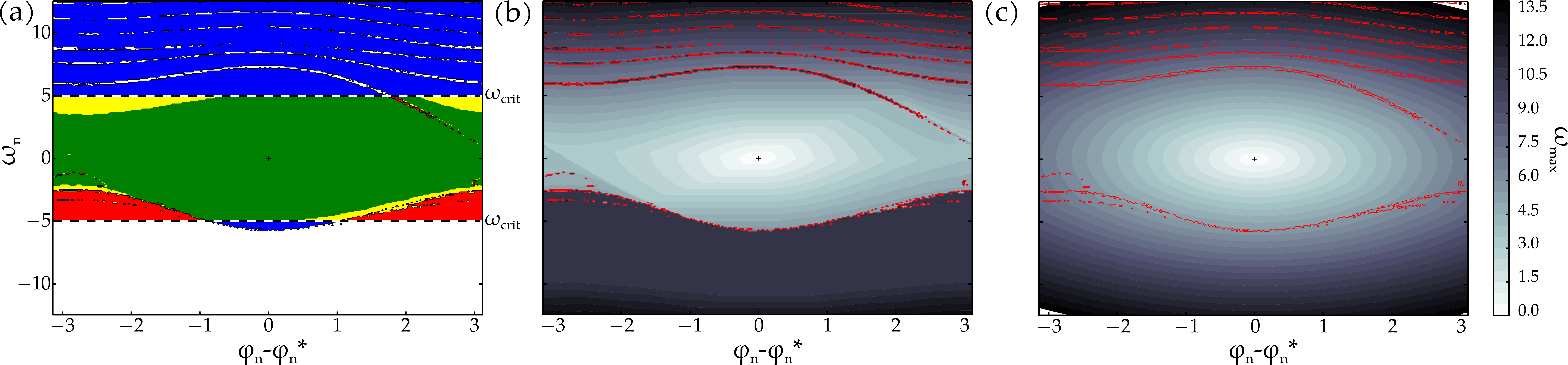}
\caption{\textbf{{Single-node phase space of a consumer in the Scandinavian grid}:} (Colour online.) 
\textbf{(a)} \edit{We plot the initial frequency deviation $\omega_n$ vs. the phase difference to the fixed point 
at node $n$, visualising the definition of the following areas using the simulation results from (b).} 
The central green area  resembles the infinite-time basin of survival, while the yellow and red areas contain finite-time 
surviving states. The union of the blue, yellow and green regions
 resembles the synchronous state's basin of attraction, while trajectories starting in the white or red regions approach different 
 attractors. The frequency threshold is chosen as $\omega_{crit.} = \pm5$ and initial conditions correspond to perturbations 
 at a single consumer node of the network. 
 \textbf{(b)} {Simulated maximum frequency deviations $\omega_{max}$} along all 
dimensions, measured over the time evolution of the system for initial conditions that correspond to perturbations at node $n$
of the network. For comparison with (a), we give the numerically estimated basin of attraction's boundaries in red.
\textbf{(c)} Corresponding analytic upper bound for the maximum
frequency deviation (cf. Eqn.\ \ref{eq:max trajectory}) for the linear approximation.}
\label{fig:northern_pspace}
\end{figure}

The basin of attraction in the conditional subspace $C_n$ of this system is illustrated in Fig.~\ref{fig:northern_pspace} (a). 
Concerning survivability, there is a subdivision in three different sets. The desirable region contains infinite- (central green region) 
and finite-time surviving states (yellow and red regions in the band).
Trajectories commencing from the remaining states within the basin of attraction (blue region) eventually reach the attractor 
asymptotically.  Note that there are also finite-time surviving states outside the basin of attraction (red region).
A large part of the single-node basin of attraction is centred around the fixed point $(\phi^*, 0)$. 
Within this region we expect the linear approximation to provide a lot of information on the system.

Regarding survivability, Fig.~\ref{fig:northern_pspace} (b) shows that the frequency deviations inside the basin of attraction do indeed 
become large. The shape of the level lines of the frequency deviations corresponds to the basins of survival for 
different frequency constraints.

Fig.~\ref{fig:northern_pspace} (c) shows the bound for the frequency deviation of the linearized dynamics calculated according to 
Eqn.~\ref{eq:max trajectory}. This shows a good qualitative agreement with the actually simulated frequency deviations 
as long as the deviations remain \edit{close to the fixed point, e.g. in the range of realistically allowed perturbations 
(see above)}. Still, the impact of the non-linearity (e.g. multistability is not captured)
on the system becomes apparent, especially further away from the fixed point.

\begin{figure}[!ht]
\centering
\includegraphics[width=\columnwidth]{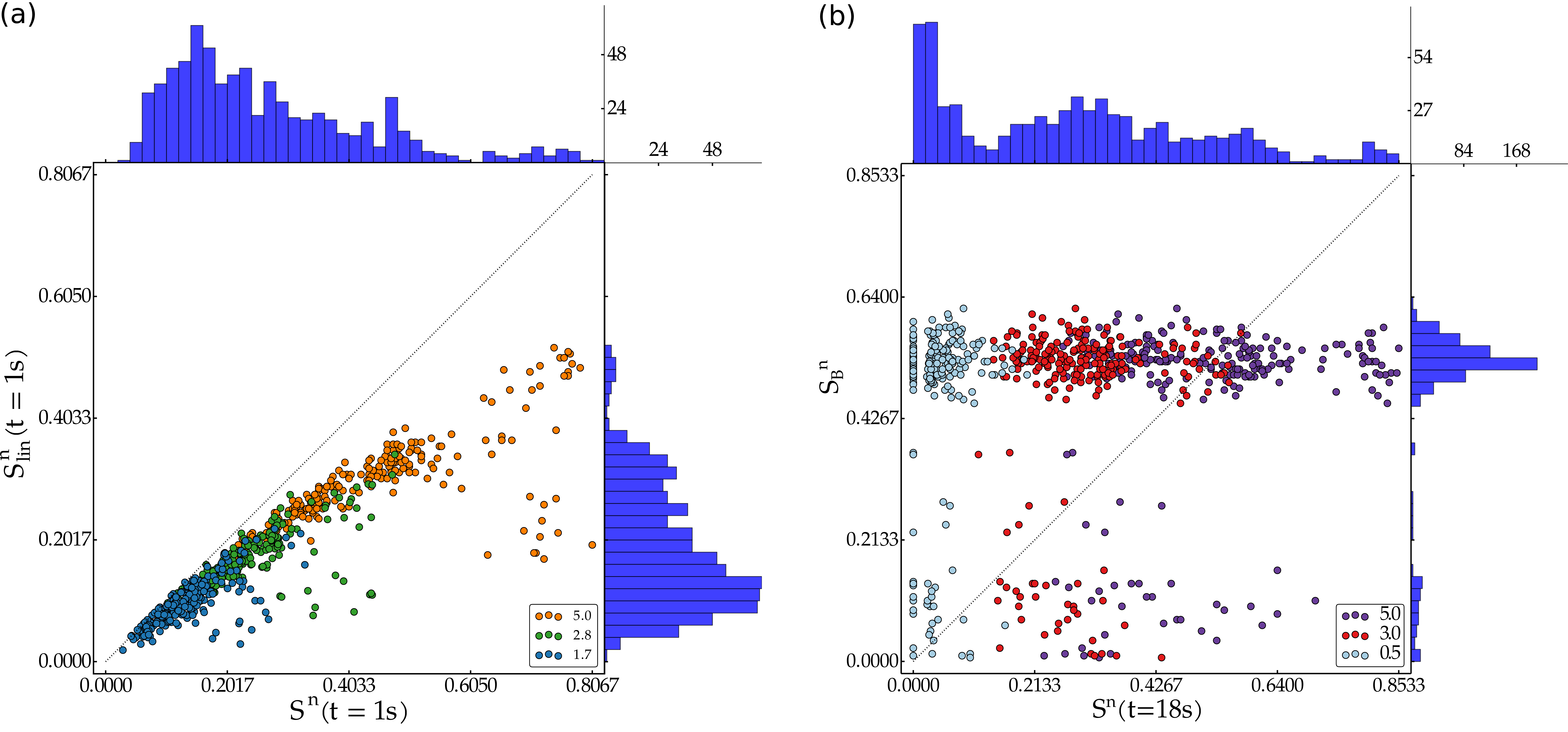}
\caption{\textbf{(a) Simulated vs. approximated single-node survivability for the Scandinavian grid:} (Colour online.) 
Scatter plot of the simulated $S^n(t)$ vs. approximated single-node survivability \edit{$S^n_{lin}(t)$} (cf. Eqn.~\ref{eq:max trajectory}) estimated for all nodes in the Scandinavian power grid ($\omega_{crit.}$ is indicated in the legend). 
The corresponding distributions are given \edit{on the sides}.\\
\textbf{(b) Single-node basin stability vs. single-node survivability for the Scandinavian grid:} (Colour online.) 
Scatter plot of the single-node basin stability $S_B^n$ vs. single-node survivability $S^n(t=100)$ ($\omega_{crit.}$ is indicated in the legend) estimated for all nodes in the Scandinavian power grid. The corresponding distributions are given \edit{on the sides}. Note that we have chosen the initial region $X_0$ for single-node basin stability with $|\omega| < 100$, the same region as in \cite{Menck2014}.}
\label{fig:bos-vs-bs}
\end{figure}

Indeed Fig.~\ref{fig:bos-vs-bs}a shows that there is a high correlation between the lower bound of the survivability of the
linear system $S^n_{lin}(t)$ calculated according to Eqn.~\ref{eq:max trajectory} \edit{(see \emph{Methods section})}
 and the actual survivability $S^n(t)$ at the 
majority of nodes for realistic 
values of frequency deviations. What exactly gives rise to the outliers far below the diagonal will require further study.
{It is important to emphasise} that the computational cost of calculating the bounds on the maximum frequency deviation
for a sample of initial conditions is many orders of magnitude lower than the numerical estimate of the survivability
via simulations of the actual time evolution. For a realistic network size of several hundred nodes, 
the {approximate} calculations can be performed on a laptop computer in less than a minute, whereas 
the {numerical survivability estimation} took several hours on 200 nodes of a computing cluster.

Fig.~\ref{fig:bos-vs-bs}b shows $S_B^n$ as well as the single-node survivability of nodes in the 
Scandinavian power grid. We see that there is no significant correlation between the two quantities.
This proves the point that the asymptotic behaviour of the system is not a strong indicator of the transient behaviour, 
at least in the case of power grids. The information we obtain from the survivability analysis is genuinely new information.

\begin{figure}[!ht]
\centering
\includegraphics[width=\columnwidth]{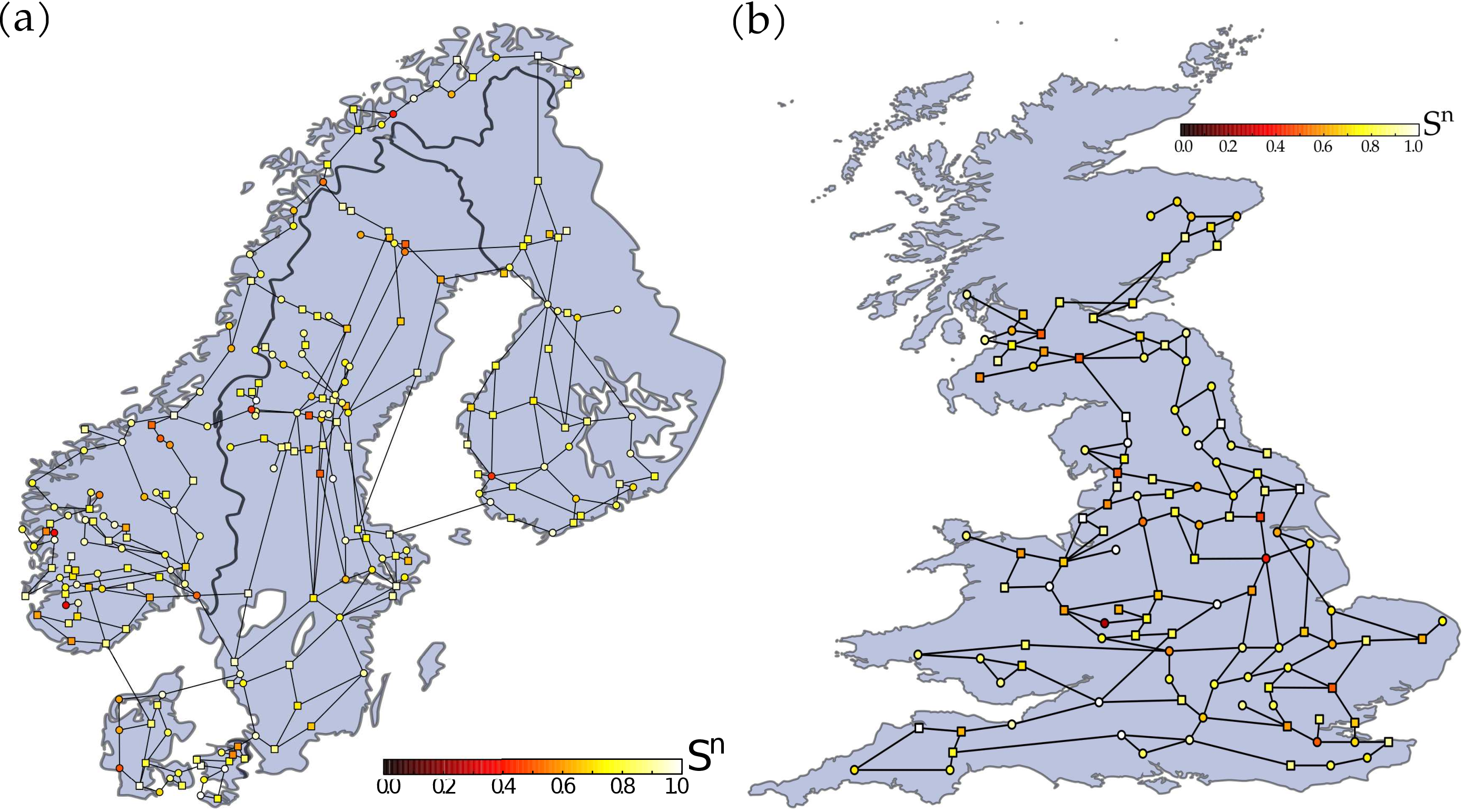}
\caption{\textbf{(a) Scandinavian power grid:} (Colour online.) The nodes' colouring indicates the respective 
single-node survivability estimate  $S^n(t=1s)$ in the Scandinavian power grid. The frequency threshold is 
chosen as $\omega_{crit.} = \pm10$.  We randomly selected a dispatch scenario, circular nodes are net generators, 
squares are net consumers. 
The map of Scandinavia has been modified from 
\burl{https://commons.wikimedia.org/wiki/File:Scandinavia.svg}, which is licensed under the 
Attribution-Share-Alike 3.0 Unported license. The license terms can be found on the following link: 
\burl{https://creativecommons.org/licenses/by-sa/3.0/}.\\ 
\textbf{(b) UK power grid:} (Colour online.) Single-node survivability estimate  $S^n(t=1s)$ of the UK power grid. 
Details analogous to (a).
The map of Great Britain has been modified from 
\burl{https://commons.wikimedia.org/wiki/File:England,_Scotland_and_Wales_within_the_UK_and_Europe.svg}, 
which is licensed under the Attribution-Share-Alike 3.0 Unported license. The license terms can be found on the following link: \burl{https://creativecommons.org/licenses/by-sa/3.0/}.}
\label{fig:grids}
\end{figure}

The Scandinavian power grid\cite{Menck2014} consists of $N=236$ nodes and $320$ links, corresponding 
to a mean degree of $\bar k = 2.7$. Hence, it has a sparse network topology with 
only a few neighbours per node on average, which is typical for power grids in general, independent from 
the number of nodes \cite{Schultz2014a}. The same holds for our second data set, the UK high-voltage transmission grid,
which consists of $N=120$ nodes and $165$ links, corresponding to a mean degree of $\bar k = 2.8$.

In Fig.~\ref{fig:grids}a and Fig.~\ref{fig:grids}b we show the geographically embedded Scandinavian and UK 
power grid. The colour of each node corresponds to the single-node conditional survivability $S^n(t=1s)$. Different nodes
exhibit starkly different {survivability} to perturbations. We find that at a threshold of $|\omega_{crit}| = 10$, for both of these realistic 
power grid topologies, there are a \edit{few nodes} that are particularly vulnerable to perturbations. This means a
perturbation at these nodes is very likely to be amplified temporarily by the overall grid dynamics. What exactly 
leads to this vulnerability, and how to characterise it in terms of grid parameters and topology is a question for future work.

Finally, we also found that the survivability in this system asymptotes very quickly. 
Simulating just the first second of the power grid is typically sufficient, the so-called ``first swing''
following a disturbance mainly determines the overall frequency deviation.

Let us summarise the key points from applying survivability to power grids:
\begin{itemize}
\item For realistic small deviations, the upper bound applied to the linear approximation provides an excellent picture 
of the infinite-time basin of survival. {The fact that the bulk of nodes shows a high correlation at large perturbations 
indicates that $S^n$ can still be determined from the approximation in this case.}
\item For the given dynamics, the survivability very quickly reaches its asymptotic value. We expect this to be a fairly generic phenomenon if we are dealing with damped systems near a stable fixed point.
\item Conditioning the survivability on regions of phase space with special meaning, like perturbations at a single node, allows  us to reveal a large amount of non-obvious structural information on a networked system. Further work is needed to understand what gives rise to the revealed structure in realistic power grids.
\end{itemize}

\section*{Discussion}

Survivability is a novel stability concept complementary to basin stability $S_B$ and linear methods of asymptotic stability analysis.
It applies to linear and non-linear systems, in the absence and presence of multi-stability. It focuses on transient rather than 
asymptotic behaviour, and incorporates exogenous information via {assuming} a desirable region for the system dynamics. 
Further, survivability can be estimated numerically at low computational costs, comparable to or even lower than for 
estimating $S_B$.

For linear systems we provide easy to evaluate analytic and semi-analytic expressions for lower bounds of the survivability, 
with a trade-off between the quality of the bound and numerical cost for evaluating the analytic expression.
These {reduce} the need to simulate the system, yielding further dramatic improvements in computational cost.

The bounds we find demonstrate that the survivability depends crucially on the eigenvectors of the linear dynamics, 
rather than the eigenvalues \edit{(see discussion in the \emph{Methods} section)}. It is an effective 
measure of the interaction between external constraints and the geometry of the dynamics in its phase space.
The fact that the bound is tight exactly when the analysis of asymptotic stability using the eigenvalues of the linearized system fails 
shows that the survivability is genuinely complementary to eigenvalue-based stability concepts.

To explore this measure in practice, we analyse three conceptual examples. 

\textbf{Carbon Cycle:} We observe that survivability accurately exhibits the presence of dangerous transient behaviour in the model, something that $S_B$ can
not detect. 
\edit{The almost monotonous decrease towards the first tipping point, opposed to the discontinuous $S_B$ curve, 
shows the potential to derive an early warning scheme from an observation of these measures for certain kinds of bifurcations.}
Just as for $S_B$, the problem of 
evaluating the survivability from data remains a challenge for future work. 

\textbf{Neuronal Networks:} Here, the transients do not arise from perturbations constructed as deviations around a desirable {attractor},
but they are randomly chosen from the whole compact phase space. Rather, 
the main interest lies on the transients themselves. Survivability reveals the same qualitative dependence of the dynamical 
behaviour on the underlying network topology as the average length of the transient\cite{Zumdieck:2004be}. 
Beyond that, considering $S(t)$ at fixed $t$ as a function of the underlying topological parameters enables us to look in more 
detail into the relationship between function and structure of pulse-coupled oscillator networks. In contrast to the average 
length of the transients, the survivability also has a direct conceptual interpretation as the probability of the system 
remaining in the interesting transient regime.
Thus it captures the appropriate notion of {stability} of transient chaos against the global attractor.

\textbf{Power Grid:} In this example we can see \edit{in} detail the interplay between the semi-analytic bounds that we 
developed and the fully non-linear system. We demonstrate that survivability under realistic constraints captures information about 
the system not contained in the $S_B$ estimate. We also demonstrate that the semi-analytic lower bounds, are strongly
correlated with the simulations of the non-linear dynamics. Thus they contain much of the relevant information about the system.
In strategic power grid development studies, this fact becomes particularly important as computational power is often at a \edit{considerable constraint}, due to the need to simulate a wide range of divergent future scenarios of the energy transition. Dynamical properties outside of 
quasi-stationary calculations can only be taken into account if efficient estimators exist, since it is not feasible to run simulations. 
Thus our lower bounds, which eliminate the need for such simulations, potentially enable a more systematic way to investigate the
 impacts of the energy transition. In particular, the influence of changing topologies and different distributions of dynamical parameters 
 on the dynamics of the power grid become computationally accessible.
For the application to power grids, there are many more operational conditions on the system's behaviour that we do not consider
here. While not all of them are as amenable to analytic considerations as the frequency deviation, we anticipate that it will still be
possible to find cheap analytic boundaries for them. The reason that we could calculate the lower bounds so easily is that the 
phase space geometry is encoded in an efficient way in the eigenvectors. This aspect will carry over to many other, more
complicated exogenous boundaries.

We thus have seen that the notion of survivability is general 
and powerful enough to capture the interplay between externalities and the intrinsic dynamics in three vastly different examples. 
\edit{In particular the last example demonstrated both the utility of single node survivability, revealing structural weaknesses and 
strengths of realistic power grid topologies, as well as of our semi-analytic bounds, reducing computational efforts dramatically.}

The work presented here thus opens up a plethora of new avenues of research. On the theoretical side, the existence of a closed form 
lower bound on the survivability of a linear system opens the door to study the survivability as a function of the network 
topology and system parameters analytically, especially for the optimisation of these parameters to increase the system's {survivability}.
The lower bounds presented here can certainly be improved by taking the more detailed geometry of the trajectories of the 
linear system into account. It will also be important to extend them to the types of bounds we have in more realistic power grid models.

\section*{Methods}

\subsection*{Numerically estimating survivability} \label{sec:numerics}

One advantage shared by survivability and  \cite{Menck2013} is that they can be efficiently estimated by randomly
sampling starting conditions. A trajectory either survives or not, therefore we can regard 
the sampling as a Bernoulli {experiment} with probability given by $S(t)$, hence 
the standard error ($SE$) of the {probability estimator} of a trial with $N$ draws is simply
\begin{equation}
 SE = \sqrt{\frac{S(t)(1-S(t))}{N}}\, .
\end{equation}
As a crucial consequence, the standard error of  a survivability estimation does not depend on the dimensionality of the system. Further, 
the condition that a trajectory has left $X^+$ tends to be easier to evaluate in practice than whether the trajectory is asymptotically 
approaching a fixed point. Furthermore, in numerical simulations, an integration might be stopped once $X^+$ has been left.

\subsection*{Analytic results for linear systems} \label{sec:analytics}

An important analytically tractable case is the total survivability $S_\infty$ for a linear dynamic in 
$X = \mathbb{R}^N$, the Lebesgue measure $\vol( X) = \int_{ X} \mathrm{d}x^N$, and a polyhedral desirable region given 
by $m$ linear conditions  $y_k \cdot x(t) < 1$ for a set of vectors
 $y_k$, $k = 1\dots m$ in $\mathbb{R}^N$. In this case we can give a lower bound on $\vol(X^S_\infty)$ that is easy to evaluate.

In this section we briefly give the results necessary for the applications in the results section on power grids. There we demonstrate that the semi-analytic bound captures the survivability of the system quite 
accurately in practical examples. In the {SI} we show detailed derivations, as well as further analytic results.

Consider a system of linear ordinary differential equations
\begin{equation}
 \dot x(t) = \mL x(t)
\end{equation}
with $x \in X=\mathbb{R}^N$ and $\mL \in \mathbb{R}^{N\times N}$ with all eigenvalues having non-positive real parts.
In general, $\mL$ has a complex spectrum. The eigenvectors $v_j$ of the complex eigenvalues are real or come in complex
conjugate pairs, from which we pick one eigenvector each. We then define the $N\times N$ matrix $\mathbb{V}$  by stacking 
the eigenvectors, or their real and imaginary parts respectively, against each other as column vectors:
\begin{equation}
\mathbb{V}  =  [v_1, \dots, \re{v_{j}}, \dots, - \im{v_{j}},\dots].
\end{equation}

\edit{This allows us to translate initial conditions into the eigenvector basis by setting $c' = \mathbb{V}^{-1} x(0)$, and combining 
$c'_{k}$ into complex numbers as appropriate $c_{j} = c'_{k} + i c'_{k + n_c}$, where $n_c$ is the number of complex eigenvalues. 
Then the trajectory describes an exponential decay along the real eigenvectors and an inward spiral in the $\re{v_{j}}$, $\im{v_{j}}$ 
plane that is parametrised by $c_{j}$, and given by $\re{\exp(\lambda_j t) c_j v_j}$. We then obtain an upper bound for the 
deviation of the trajectory starting at $x(0)$ in a direction $y_k$ 
by maximising the contribution of each eigenvector separately.}

Now, setting $y_{kj} := y_k \cdot v_j$ for $v_j$ real, and $y_{kj} := |y_k \cdot v_j|$ for $v_j$ complex, this leads to the estimate:

\begin{equation}
\label{eq:max trajectory}
\mmax_{t\in[0;\infty[} |y_k \cdot x(t)| \leq  \sum_{j=1}^{n_0} y_{kj} c_j +
 \sum_{j=n_0 + 1}^{n_r} \max(0, y_{kj} c_j) 
+ \sum_{j=n_r + 1}^{n} y_{kj} |c_j|
\end{equation}

where the first sum is over real eigenvectors corresponding to null eigenvalues, the second is over nonzero real eigenvectors and 
the last is over the complex eigenvectors.

Setting the right hand side of Eqn.~\ref{eq:max trajectory} smaller than $1$ defines a region
$V_c$ in $\mathbb{R}^N$ spanned by the real and imaginary parts of the coefficients $c_j$. 
\edit{This region is mapped to the state space by $\mathbb{V}$ and thus its volume is related to the corresponding region in phase space by a determinant factor. As it is defined by a weaker inequality than $X^S_\infty$ it follows that}

\begin{equation}\label{eq:bound-with-Vc}
 \vol(X^S_\infty) \geq \sqrt{\det{\mathbb{V}\mathbb{V}^T}} \vol(V_c)\; .
\end{equation}

The inequalities Eqn.~\ref{eq:max trajectory} together with the matrix $\mathbb{V}$ can be used to efficiently estimate the total
 survivability as well as the conditional survivability. Remarkably, for systems with a purely imaginary spectrum, the bounds of Eqn.~\ref{eq:max trajectory} and Eqn.~\ref{eq:bound-with-Vc} hold with equality.

In the {SI} we also derive a lower bound for $\vol(V_c)$.

This lower bound demonstrates that for the survivability of a linear system, the eigenvectors play a crucial role. In fact, the 
eigenvalues do not enter the bound at all, except in terms of classifying the corresponding eigenvectors in separate classes. This 
demonstrates that the survivability captures substantially different information about the linear system than eigenvalue-based {stability}
measures like relaxation time, or the master stability function.

\subsection*{Relationship to Similar Concepts}

Survivability is related to a number of concepts in other fields, notably control theory. From this perspective it can be seen as 
a so far unstudied, simplifying case where a number of distinct concepts from various fields intersect. In this section we discuss 
a number of such concepts and their precise relationship to survivability.

Survivability is conceptually similar to the notion of finite time stability as studied for linear control 
systems\cite{abdallah2003necessary,amato2001finite}.
There the focus is on finding a particular control scheme that will ensure that the resulting closed loop system stays within 
a particular region for some time, possibly in the presence of perturbations of the dynamical equations.
From our perspective this can be seen as attempting to find systems with $S(t) = 1$. As the focus there is on perturbed 
dynamics in linear control systems, the actual overlap of methods is very small, in particular it is not possible to extend the 
methods to high-dimensional non-linear systems.

Another concept from control theory which is similar to the basin of survival is the viability kernel defined by Aubin et.\ al. 
in the context of viability theory \cite{Aubin2001, Aubin2011}. They introduce the notion of an environment $K$ that 
contains all desirable states. Within the environment, there is the so-called viability kernel $V$ \cite{Bonneuil2006, Maidens2013} 
as the set of all initial conditions from which the system \emph{can} stay within the environment. This basically is a more 
general version of our infinite-time basin of survival for non-deterministic systems or systems with multiple evolution paths 
and a management process. Consequently, $K \smallsetminus V$ corresponds to the set of finite-time surviving states in 
deterministic systems. The viability kernel's volume is proposed as a measure of the degree of viability \cite{Aubin2011}, 
in the limit of no control it thus reduces to our total survivability. However, we are not aware of this special case ever being 
considered in the context of viability theory. Whereas survivability measures the ability of the intrinsic dynamics to 
withstand perturbation, viability theory is concerned with the question of the power of control. Beyond this conceptual difference, 
evaluating survivability also requires very different technical methods, analytically as well as numerically. As far as we are aware, 
sampling based methods, which are efficient and natural for survivability, are impossible for viability. This is due to the fact that 
whether a particular point belongs to the viable set depends on the optimal control, which might not be known.

There are two concepts that share some formal similarity to survivability in the context of deterministic systems,
 transient times and open systems.

The study of transient life times\cite{Houghton2010,Lai2011,Politi2010,Rosin2014} is only related to the survivability in the 
 non-typical special case that the attractor (or a small epsilon environment around it) is the only undesirable region.
 In our example of integrate-and-fire neurons this is the case, but in the power grid there is no clear relationship 
 between the strength of the transient (which might kill the system) and the return time to the attractor. 
 In fact, there, the attractor we start from is in the desirable region. Transients life times are a special case, and not 
 a typical one, of survivability. The latter is far more general, going beyond the focus on the length of transients 
 and their distribution, and typically captures genuinely different information of the system (e.g. the linear analysis 
 mainly depends on eigenvectors, not eigenvalues).
 
The theory of open systems, on the other hand, is generally concerned with ergodic systems.
For leaky chaotic systems\cite{Altmann2013} the asymptotic behaviour of the survival probability is the key observable. 
At the formal level there is an analogy 
to our definitions, however, the total survivability, the size of the total phase space that leaks, is never considered as an observable in 
the literature. \edit{Indeed it is often the case that it is the whole phase space.} Nor is the cumulative leakage ever interpreted as a stability measure or are efficient methods to
 estimate it for high-dimensional systems being discussed. In fact, as in the case of transient times, leaky systems can be seen as a special case of our discussion. Specifically it is the conditional survivability with the conditional space chosen as the space of surviving states $X^S$.

The closest analogy to our deterministic survivability is simply the survival analysis in the the context of stochastic systems. The concept of the so-called first hitting time and survival probability \cite{Anishchenko2006, Ebeling2005,Redner2002}, \edit{which can be studied for the case of stochastic perturbations to deterministic systems by quasi-potentials \cite{Freidlin2012,Graham1984,Kraut2003}}, map directly to our work. The first hitting time $t$ measures when a system is 
expected to first hit the forbidden region $X^-$. The cumulative of the probability of first hitting the undesirable region before $t$ is then $1 - S(t)$.
 Our definitions given above can be seen as a deterministic version of these concepts. The role of stochasticity in 
 the evolution is replaced by a probabilistic initial perturbation. Here similar sampling based methods are possible and necessary. The type of semi-analytic analysis we performed for the linear case would however be hard to duplicate. From this perspective what we have demonstrated is how to successfully apply methods and concepts from stochastic systems in the study of their deterministic counterparts.

The key insight in our work, as it is for $S_B$, is that restricting ourselves to probabilistic notions enables a considerably wider applicability of our analysis, as well as new numerical and analytic methods. Put differently, by asking not about the geometry of sets in phase space but merely about their volume, we can access high-dimensional non-linear systems that are out of reach for detailed geometric analysis. The challenge then lies in defining interesting sets that capture concepts of interest. As such we take it as a confirmation for the wide interest of the specific sets that survivability is based on, that it occurs a the intersection of a number of well studied concepts.

\section*{Acknowledgements}

 We acknowledge gratefully the support of BMBF, CoNDyNet, FK. 03SF0472A, of the EIT Climate-KIC project SWIPO 
 and Nora Molkenthin for illustrating our illustration of the concept of survivability using penguins. We thank Martin Rohden for 
 providing us with the UK high-voltage transmission grid topology and Yang Tang for very useful discussions.
 
\section*{Author contributions statement}

F.H. and P.S. designed the study; F.H., P.S. and C.G prepared the data; F.H. and P.S. carried out the analysis
and prepared the manuscript. All authors discussed the results and contributed to editing
the manuscript. J.H. and J.K. supervised the study.

\section*{Additional information}

\textbf{Competing financial interests}  The authors declare no competing financial interests.


\begingroup
\renewcommand{\addcontentsline}[3]{}
\renewcommand{\section}[2]{}

\endgroup

\clearpage
\appendix

\renewcommand*{\contentsname}{Appendix}
\tableofcontents

\section{Formal Definition and Derivation of Analytic Bounds} \label{apx:bounds}
As noted in the main text, the survivability of a linear system is amenable to analytic study. In this appendix we 
will give a more mathematically precise definition of survivability and then a detailed derivation of the results 
used in the main body of the text as well as some closely related ones.

\subsection{Formal definition and basic properties} \label{sec:definitions}

Consider a dynamical system with states $x$ in a state space $X$ giving rise to trajectories $x(t)$ under 
some evolution map $\sigma(t)$.
Now we define a desirable region $X^+ \subset X$ with its complement $X^- = X \smallsetminus X^+$, where the 
former contains all states $x$ that are stated to be desirable. In the penguin example (Fig.~1)
$X^-$ would contain the cliff and the valley.
In the context of Earth System science, such a desirable region has variously been called the \emph{safe operating space} 
within \emph{planetary boundaries} \cite{Rockstroem2009} or the \emph{sunny region} \cite{Heitzig2016}.

According to the main text, we define the finite-time survivability $S(t)$ of the dynamical system
 at time $t$ to be the fraction of trajectories
 starting in $X^+$ that stay within $X^+$ for the entire duration $[0,t]$. Put another way, if entering the 
 region $X^-$ terminates the system, $S(t)$ is the fraction of trajectories starting in $X^+$ still alive after 
 time $t$. We call the part of $X^+$ from which trajectories start that stay alive at least for time $t$ the \emph{finite-time}
 or \emph{$t$-time basin of survival} $X^S_t$. We then have

\begin{equation}
 S(t) :=  S_{\mu^+}(t) = \frac{\mu(X^S_t)}{\mu(X^+)}\, ,
\end{equation}

where $\mu$ is an inner measure on $X$ determining the volume of the sets $X^S_t$ and $X^+$ in the phase space.
By construction, $S_{\mu^+}(t)$ takes values on the unit interval.

We define the total survivability $S_{\mu^+}(t \rightarrow \infty)$ as the limit 

\begin{equation}
S_\infty := \lim_{t \rightarrow \infty} S(t)\, . \label{eq:Slimes}
\end{equation}

Each $t$-time basin of survival is a subset of the previous ones 
$ X^S_t \supset X^S_{t'}$ (for $ t' > t$), as trajectories returning to
$X^+$ after leaving it once do not contribute to $X_{t'}^S$. Hence, $S(t)$ 
is monotonically decreasing and bounded by $0$ from below, therefore the limit in Eqn.~\ref{eq:Slimes} exists. The use of an inner 
measure here avoids subtleties involving non-measurable sets, like fractal \cite{Nusse1996, Nusse1996a} or riddled \cite{Alexander1992, Lai1995} basins of attraction. 

Note, however, that if $X^+$ is an open set, and the map $\sigma(t): X \rightarrow X$ is continuous for all $t$, then the images of $X^+$ 
under $\sigma(t)^{-1}$ are also open. As $X^S_t = \bigcup_{0 < t' < t} \sigma(t')^{-1} X^+$ is a union of open sets, it is itself open and 
therefore  measurable if $\mu$ is a Borel measure. Thus in this important special case, which covers all applications we are considering
 in the main text, no such subtleties exist.


In some applications, the choice for the set $X^+$ might have an infinite volume, even though $X^S_t$ becomes finite 
for sufficiently large $t$. In that case one can still consider the unnormalised measure $\mu(X^S_t)$ as a relative measure of the survivability of a system.

\subsection{Conditional survivability}

The conditional survivability $S^C(t)$ measures the response of the system to restricted perturbations. For example, we might be interested in the survivability, given perturbations that are localised at a node in a network. Given a subset of the state space $C \subset X$, we define the conditional survivability as the fraction of trajectories starting in $X^+ \cap C$ that stay in $X^+$. That is

\begin{equation}\label{eq:conditional-survivability}
S^C(t) = \frac{\mu|_C(X^S_t \cap C)}{\mu|_C(X^+ \cap C)}\; , 
\end{equation}

\noindent where $\mu|_C$ is an inner measure on the smallest sub-manifold containing $C$. In the case of a network and perturbations at a single node, the phase space typically is the product of phase spaces at the nodes, and the volume measure likewise factorises, thus there are natural choices for $C$ and $\mu|_C$.

\subsection{Linear Systems}

We consider the case of a linear dynamic in $X = \mathbb{R}^N$, the standard Lebesgue measure 
$\vol( X)=\int_{ X} \mathrm{d}x^N$ and a polyhedral sunny region given by $m$ linear conditions $y_k \cdot x(t) < 1$
for a set of vectors $y_k$, $k = 1\dots m$ in $\mathbb{R}^N$.

The dynamics is given by a linear system of ordinary differential equations

\begin{equation}
 \dot x(t) = \mL x(t)
\end{equation}

with $x \in X=\mathbb{R}^N$ and $L \in \mathbb{R}^{N\times N}$.
In general, $\mL$ has a complex spectrum, and we assume that all eigenvalues have non-positive real part. We denote the number of real eigenvalues of $\mL$ as $n_r$, the number of pairs of
 complex conjugate complex eigenvalues $n_c$. Then we have a total of $n = n_r + n_c$ independent eigenvalues and 
 $n_r + 2  n_c = N$. We denote the eigenvalues as $\lambda_i$ and $\overline{\lambda_i}$, and the corresponding eigenvectors 
 as $v_i$ and $\overline{v_i}$ respectively.

Assuming that the real and imaginary parts of the eigenvectors of $\mL$ span the entire space $X$, the general solution to the 
dynamical equations are then given by the matrix exponential

\begin{align}
x(t) &= \exp{\mL t}x(0)  \nonumber\\
 &= \sum\limits_{j=1}^{n} \re{c_j\exp{\lambda_jt}v_j} \nonumber\\
 &= \sum\limits_{j=1}^{n_r} c_j \exp{\lambda_jt}v_j + \sum\limits_{j=n_r + 1}^{n}\re{ c_j\exp{\lambda_jt}v_j} \, ,
\end{align}

where the coefficients $c_j$ are real for $j \leq n_r$ and complex above. To determine them we introduce a convenience map $\iota$ as follows:

\begin{align}
\iota&: \mathbb{R}^{n_r}\otimes\mathbb{C}^{n_c} \rightarrow \mathbb{R}^N\nonumber\\
\iota&(c)_j = \begin{cases} c_j &\mbox{if } j \leq n_r \\ \re{c_j} &\mbox{if } n_r < j \leq n_c \\ -\im{c_{(j - n_c)}} \; &\mbox{if } n_c < j\; . \end{cases}
\end{align}

This is a real-linear map. Then we can define the real matrix $\mathbb{V}$ as:

\begin{align}
\nonumber \mathbb{V}  = & [v_1, \dots,v_{n_r}, \\
\nonumber & \re{v_{n_r + 1}}, \dots,\re{v_{n}}, \\
- & \im{v_{n_r + 1}},\dots, -\im{v_{n}}].
\end{align}

and obtain

\begin{equation}
x(0) = \br{\mathbb{V}\circ \iota}\br{c}\, ,
\end{equation}

which can be readily inverted as we assumed $\mathbb{V}$ to have full rank.

\subsection{Upper bound on the deviation of a single trajectory.}

The inner product of $x(t)$ with a boundary vector $y\in\mathbb{R}^N$ is then simply given by:

\begin{equation}
y \cdot x(t) = \sum_{j=1}^{n_r} c_j\exp{\lambda_jt} y \cdot v_j + \sum_{j=n_r+1}^{n_r+n_c} \re{c_j\exp{\lambda_jt} y \cdot v_j}
\end{equation}

Now, to obtain an upper bound on this for all times we can maximise each individual contribution. As the eigenvalues have non-positive real part
each complex contribution has magnitude of at most $|c_j y \cdot v_j|$. The maximum of the real contribution depends on the 
sign of $c_j y \cdot v_j $ and is given by \begin{equation}\max(0, c_j y \cdot v_j)\end{equation} unless $\lambda_j = 0$ in 
which case the contribution is exactly equal $c_j y \cdot v_j $. While this does not happen generically it occurs due to symmetries 
in the system, and thus we will treat it separately. We denote the number of zero eigenvalues by $n_0$. We have:

\begin{align}\label{eq:general bound}
\mmax_{t\in[0;\infty[}|y \cdot x(t)| \leq& \sum_{j=1}^{n_0} c_j y \cdot v_j \nonumber\\+& \sum_{j=n_0 + 1}^{n_r} \max(0, c_j y \cdot v_j) \nonumber\\+& \sum_{j=n_r+1}^{n} |c_j y \cdot v_j| \; .
\end{align}

This is the key approximation for our analysis. In the next section we will use this estimate to give a lower bound for the total 
survivability, afterwards we will show when this bound becomes tight.

\subsection{A lower bound for the total survivability}

For a boundary vector $y_k$ let us define $y_{kj} = y_k \cdot v_j$  for $j \leq n_r$ and $ y_{kj} = |y_k \cdot v_j|$ for $n_r < j \leq n$. 
Then the inequalities
\begin{equation}\label{eq:Vc}
\sum_{j=1}^{n_0} y_{kj} c_j + \sum_{j=n_0 + 1}^{n_r} \max(0, y_{kj} c_j) + \sum_{j=n_r + 1}^{n} y_{kj} |c_j|  < 1
\end{equation}
define a region $V_c$ in $\mathbb{R}^{n_r}\otimes \mathbb{C}^{n_c}$ that is mapped to a subset of $X^S_\infty$ by the linear 
transformation $\mathbb{V} \circ \iota$. This is a subset of $X^S_\infty$ as the initial conditions in this region have an inner 
product with the boundary vectors $y_k$ bounded by Eqn.~\ref{eq:general bound}. Thus, taking the effect of the transformation 
$\mathbb{V}$ into account, we have the lower bound

\begin{equation}\label{eq:Vc bound}
 \vol(X^S_\infty) \geq \sqrt{\det{\mathbb{V}\mathbb{V}^T}} \vol(V_c)\; .
\end{equation}

This bound can be evaluated numerically quite easily. All that is needed is to sample $X^+$ and take the fraction of samples for which 
the image under the linear map $(\mathbb{V} \circ \iota)^{-1}$ satisfies Eqn.~\ref{eq:Vc}. This also makes it very easy to
numerically estimate the lower bound of conditional survivability, by sampling from $C \cap X^+$ instead. The data for the semi-analytic figures in the power grid section of the results was computed in this way.

In order to proceed with the analytic calculations we have to consider further special cases. First we will show that the bound Eqn.~\ref{eq:Vc bound} is actually exact for some cases.

\subsection{The case of vanishing real parts.}

Let us now consider the case where the real part of all $\lambda_i$ is zero, and $\frac{\im{\lambda_i}}{\im{\lambda_j}}$ is irrational. Thus $n_0 = n_r$.
In that case the trajectory with initial conditions $x(0) = \mathbb{V} \circ \iota \circ c$ is dense on the torus:

\begin{equation}
\mathbb{T} = \left\{\sum_{j=1}^{n_r} c_j v_j + \re{\sum_{j=n_r + 1}^{n} c_j \exp{i \phi_j} v_j}\middle\vert \phi_j \in [0;2\pi[\right\}
\end{equation}
 
The maximum of the torus along the $y$ direction is obtained exactly by maximising each contribution independently, which means tuning the $\phi_i$ so that $c_j\exp{i \phi_j} \hat{x} \cdot v_j$ are real, thus the real part equals the absolute value:

\begin{equation}
\mmax_{\phi_i\in[0; 2\pi[} |y \cdot \mathbb{T}|\leq \sum_{j=1}^{n_r} c_j y \cdot v_j + \sum_{j=n_r + 1}^{n} |c_j y \cdot v_j| 
\end{equation}

with equality if $\frac{\im{\lambda_i}}{\im{\lambda_j}}$ is irrational. Therefore, in this case, the general bound 
Eqn.~\ref{eq:general bound} holds with equality. We have:

\begin{equation}
\mmax_{t\in[0;\infty[} |y \cdot x(t)| = \sum_{j=1}^{n_r} c_j(x_0) y \cdot v_j + \sum_{j=n_r+1}^n |c_j(x_0)| |y \cdot v_j| 
\end{equation}

and therefore also

\begin{equation}
 \vol(X^S_\infty) = \sqrt{\det{\mathbb{V}\mathbb{V}^T}} \vol(V_c)\;.
\end{equation}

\subsection{The purely imaginary case.}

For the case $n_r = n_0 = 0$ we can give an explicit lower bound for $\vol(V_c)$. To do so, define $\tilde y_j = \mmin_k y_{kj}$. The volume of the space $\tilde V_c$ defined by the inequality

\begin{equation}
\sum_{j=1}^{n} \tilde y_{j} |c_j|  < 1
\end{equation}

is a lower bound for the volume of $V_c$. The two spaces have the same volume if there is one condition that dominates all others, that is, if there is a $k'$ such that $\tilde y_j = y_{k'j}$ for all $j$. This means that any trajectory that leaves the allowed region also crosses the boundary defined by $x \cdot y_k \leq 1$.

We now need to evaluate the $2n = 2n_c = N$ dimensional integral

\begin{equation}
\vol(\tilde V_c) = \int_{\tilde V_c} \prod_{j=1}^n \mathrm{d} c^r_j \mathrm{d} c^i_j\;,
\end{equation}

with $c_j = c^r + i c^i_j$. We begin by changing $\mathrm{d} c^r_j \mathrm{d} c^i_j$ to polar coordinates $r_j \mathrm{d} r_j \mathrm{d} \phi_j$ and rescaling:

\begin{align}
\nonumber \vol(\tilde V_c) & = \left(\prod_j \tilde y_j^2\right) \int_{\sum_j r_j < 1} \prod_{j=1}^n r_j \mathrm{d} r_j \mathrm{d} \phi_j \\
 & = (2 \pi)^n \int_{\sum_j r_j < 1} \prod_{j=1}^n r_j \mathrm{d} r_j\;.
\end{align}

The  integral can now be written as

\begin{equation}
\vol(\tilde V_c) = \left(\prod_j \tilde y_j^2\right) (2 \pi)^n \int_0^{1 - \sum_{j > 1} r_j} r_1 \mathrm{d} r_1 \int_0^{1 - \sum_{j > 
2} r_j} r_2 \mathrm{d} r_2\int_0^{1 - \sum_{j > 3} r_j} r_3 \mathrm{d} r_3\dots \;.
\end{equation}

This can be calculated by beta functions\footnote{The detailed calculation can be found here \href{http://math.stackexchange.com/a/207605}{http://math.stackexchange.com/a/207605}. 
(Accessed: \today)}:

\begin{equation}
\vol(\tilde V_c) = \frac{ (2 \pi)^n}{(2n + 1)!} \prod_j \tilde y_j^2\; .
\end{equation}

This finally means that we obtain the lower bound on the region of total survivability of a linear system with no real eigenvalues and all real parts of the eigenvalues equal to zero of

\begin{equation}
 \vol(X^S_\infty) \geq \frac{ (2 \pi)^n}{(2n + 1)!} \sqrt{\det{\mathbb{V}\mathbb{V}^T}} \prod_j \tilde y_j^2\; .
\end{equation}

\section{Relationship to Basin Stability} \label{apx:basin}

Let us further consider the relationship between basin stability and survivability.
Consider the union of all attractors $A$ in $X$. Following the terminology of \cite{Heitzig2016}, we split the set $A$ 
 into desirable attractors $A^+$ and undesirable attractors $A^-$. Define $X_A^+$ and $X_A^-$ to be the basin of attraction of 
 $A^+$ and $A^-$ respectively. The basin 
stability of $A^+$ with respect to some initial region $X^0$ is then defined as
\begin{equation}
 S_B = \frac{\vol(X^0 \cap X_A^+)}{\vol(X^0)}\, .
\end{equation}

Note, that defining basin stability requires knowledge of the respective attractor. Efficiently evaluating it numerically 
requires a criterion to evaluate whether the system will converge to a certain attractor.

Let us assume that the sunny region cleanly separates the set of attractors, that is, there are some attractors that are 
entirely sunny and others that are entirely shaded, but none that intersect both regions.
In this case, we can establish a quantitative relationship between basin stability and survivability. 
We can choose $A^+ = A \cap X^+$ (i.e. desirable attractors are contained inside the sunny region), and we know that 
asymptotically the trajectories converge either to $A^-$ or $A^+$ or diverge. Thus every trajectory that does not 
contribute to the basin stability also has to leave the sunny region eventually and can not contribute to the survivability either. 
Thus if

\begin{align}
&& A^+ &= A \cap X^+\nonumber\\
\text{and}&& X^0 &= X^+\, ,\nonumber\\
\text{then}&& S_B & \geq S_\infty\, . 
\end{align}

The difference between the two values is exactly the measure of initial conditions whose trajectories  leave the sunny region
intermittently but eventually return to it and stay. Thus we see that whether basin stability or survivability is the appropriate
measure depends on whether the forbidden region is merely unpleasant, and we want our stay there to be finite, or whether the
forbidden region is deadly and we absolutely do not want the system to enter it at all.

\section{Pulse-Coupled Integrate-and-Fire Oscillators} \label{sec:neuron-model}

We here give further details about the integrate-and-fire model for coupled neurons we used in the main text.
Firstly, the free dynamics of an oscillator $j$ is given by
\begin{equation}
\dot\phi_{j}(t)=1\,.\label{eq:timeevol}
\end{equation}

When an oscillator $j$ reaches the threshold, $\phi_{j}(t)=1$,
its phase is reset to zero, $\phi_{j}(t^{+})=0$, and the oscillator
emits a pulse that is sent to all oscillators $i$ possessing an in-link
from $j$. After a delay time $\tau$ this pulse induces a phase jump
in the receiving oscillator $i$ according to

\begin{equation}
\phi_{i}((t+\tau)^{+}):=\min\left(1, \frac{\exp{b\varepsilon_{ij}} - 1}{\exp{b} - 1} + \exp{b\varepsilon_{ij}}\phi_{i}(t+\tau)\right) \label{eq:phasejump}
\end{equation}

The phase dependence is determined
by a twice continuously differentiable function $U(\phi)$ that is
assumed to be strictly increasing, $U'(\phi)>0$, concave (down),
$U''(\phi)<0$, and normalised such that $U(0)=0$ and $U(1)=1$.

This model, originally introduced by Mirollo and Strogatz \cite{MIROLLO:1990p366},
is equivalent to different well known models of interacting threshold
elements if $U(\phi)$ is chosen appropriately. Here we take functions of the form

\begin{equation}
U_{b}(\phi)=b^{-1}\ln(1+(e^{b}-1)\phi),
\end{equation} 

where $b>0$ parametrises the curvature of $U$, that determines
the strength of the dissipation of individual oscillators. The function
$U$ approaches the linear, non-leaky case in the limit $\lim_{b\rightarrow0}U_{b}(\phi)=\phi$.
Other nonlinear choices of $U\neq U_{b}$ give results similar to
those reported above. 

The considered graphs are strongly connected, i.e. there exists a directed path between any pair of nodes.
We normalise the total input to each node $\sum_{j=1}^{N}\varepsilon_{ij}=\varepsilon$ such that the fully 
synchronous state exists. Furthermore for any node $i$
all its $k_{i}$ incoming links have the same strength $\varepsilon_{ij}=\varepsilon/k_{i}$.

\begingroup
\renewcommand{\addcontentsline}[3]{}
\renewcommand{\section}[2]{}

\endgroup

\end{document}